\def\gdot{\dot\gamma}
\def\nism{n_{\rm ISM}}
\def\nismo{n_{\rm ISM, 0}}
\def\npair{n'_{\rm pair}}
\def\Thpair{\Theta_{\rm pair}}
\def\ee{$e^-e^-$}
\def\ep{$e^-e^+$}
\def\ls{\lower4pt\hbox{${\buildrel < \over \sim}$}}
\def\gs{\lower4pt\hbox{${\buildrel > \over \sim}$}}

\documentclass[12pt, preprint]{aastex}

\slugcomment{Submitted to {\it The Astrophysical Journal} on June 5, 2001\\
Revised: July 24, 2001}

\shorttitle{collision-dominated pair fireballs}
\shortauthors{B\"ottcher et al.}

\begin{document}

\title{Radiation from collision-dominated relativistic pair fireballs}

\author{M. B\"ottcher\footnote{Chandra Fellow}}
\affil{Department of Physics and Astronomy, Rice University, MS 108, \\
6100 S. Man Street, Houston, TX 77005 - 1892, USA}
\email{mboett@spacsun.rice.edu}

\author{R. Schlickeiser}
\affil{Institut f\"ur Theoretische Physik, Lehrstuhl IV \\
Ruhr-Universit\"at Bochum, D-44780 Bochum, Germany}
\email{rsch@egal.tp4.ruhr-uni-bochum.de}

\and

\author{A. Marra\footnote{Current address: MIT, School of Engineering;
77 Massachusetts Avenue; Cambridge, MA 02139-4307}}
\affil{Stephen F. Austin High School, 3434 Pheasant Creek Drive, \\
Sugar Land, TX 77478, USA}
\email{anibal.marra@cognitas.com}

\begin{abstract}
It is generally accepted that gamma-ray bursts (GRBs) are initiated 
by a relativistic pair fireball, converting its internal energy into
kinetic energy of a relativistically moving plasmoid and subsequently
into radiation. Here, we investigate the early stages of this evolution,
after the pair fireball has become optically thin to $\gamma\gamma$ 
pair production. We show that for a short period of time, $\sim 0.1$ 
-- a few seconds after the initial explosion, the pair plasmoid 
evolution might be dominated by collisional processes prior to 
the formation of a collisionless shock. We simulate these processes 
during the early pair plasmoid evolution and calculate the expected 
radiative signatures. We show that the radiation from the
collision-dominated pair plasmoid phase results in a short 
($\sim$~a few ms) flash of thermal soft X-ray emission, followed
by a transition phase of $\ls 1$~s during which the fireball
turns Thomson thin, but its radiation remains dominated by thermal
Comptonization, peaking at around $E_{\rm pk} \sim$~100 MeV -- a few
GeV. While the very early thermal emission could be associated with 
the quasi-thermal radiation signatures found in the very early phases 
of several bright BATSE GRBs, the predicted subsequent flash of high-energy 
emission should be easily detectable with the GLAST satellite.
\end{abstract}


\section{Introduction}

With the establishment of the cosmological distance scale of
$\gamma$-ray bursts (GRBs) and the considerable success of 
the synchrotron-shock model \citep{mr93,katz94,tavani96} to 
explain the broadband (radio through X-ray) continuum afterglow 
signatures of GRBs \citep{wrm97,vietri97,waxman97,galama98,mrw98}, 
it is now generally accepted that GRBs are initiated by 
a relativistic pair fireball, transferring its internal 
energy into kinetic energy of a relativistic blast wave 
\citep{cr78,sp90,mr92,rm92}. Non-thermal particle acceleration 
at the forward shock and subsequent radiative cooling is 
believed to lead to the formation of a broken power-law 
distribution of ultrarelativistic electrons, producing the 
observed broadband afterglow radiation primarily through 
synchrotron emission (e.g., \cite{spn98,bd00}), though Compton
upscattering of synchrotron photons may also play a significant 
role \citep{dbc00,se01}. 

In the relativistic fireball/blastwave model for GRBs, a large
amount of energy is released in a short period of time. This
initial configuration is highly opaque to $\gamma\gamma$ 
absorption and pair production and leads to the formation
of a relativistic pair fireball. As the fireball expands,
it cools adiabatically, converting its internal energy into
bulk kinetic energy of the outflow, and its $\gamma\gamma$ 
opacity decreases until it becomes optically thin to 
$\gamma\gamma$ pair production. At this time, the pairs 
are essentially cold in a reference frame co-moving with a 
small section of the expanding pair plasma / radiation shell. 
It has been recognized very early-on (e.g., \cite{sp90,mr92})
that even a small intrinsic or external contamination of
the pair fireball with baryons would lead to re-conversion
of most of the fireball energy into kinetic energy of the 
outflow, and that any radiation escaping the pair fireball
during this phase of energy conversion would have quasi-thermal
signatures. 

However, more detailed measurements of time-averaged photon 
spectra emerging from GRBs, in particular by the Burst and 
Transient Source Experiment (BATSE) on board the {\it Compton 
Gamma-Ray Observatory} (CGRO) have shown that they can generally
be fitted successfully with a model consisting of two power-laws,
smoothly connected by an exponential turn-over \citep{band93}. 
Such a power-law high-energy spectrum was interpreted as the
radiative signature of non-thermal particle acceleration at
relativistic shocks formed in the process of deceleration 
of the relativistic blast wave produced during the initial
fireball stage (e.g., \cite{pr93,mr93}). While it is now
generally agreed that the smoothly decaying afterglow emission 
observed from many GRBs is produced by relativistic electrons
accelerated at the external forward shock, in the stage when 
the blast wave is efficiently being decelerated by the sweeping-up
of external material, many researchers now believe that the
emission during the prompt GRB phase may be produced in internal
shocks during collisions of subsequent relativistic blast waves
produced in a series of energy release events from the central 
engine (e.g., \cite{fmn96,sp97}), although an external-shock 
scenario for the prompt GRB phase cannot be ruled out at this 
time (e.g., \cite{dm99}). 

Both internal and external shock scenarios for GRBs are generally 
starting out with the assumption that the relativistic blast wave 
has produced a relativistic shock with fully developed hydromagnetic 
turbulence in order to allow for efficient non-thermal particle 
acceleration at the shock front. This neglects the fact that it 
takes a finite amount of time to build up the necessary turbulence 
in the two-stream multi-fluid system consisting of the relativistically 
moving pair plasma and the background hydrogen plasma. While a 
detailed treatment of the process of the development of hydromagnetic
turbulence through a relativistic two-stream instability is rather
cumbersome (see, e.g., \cite{ps00,sd00,schlickeiser01}), 
the time scale for the development of such turbulence can be 
estimated by the Alfv\'en crossing time through the shell of
relativistically moving pair plasma. Using hydrodynamics
simulations, \cite{mlr93} have demonstrated that a relativistic 
pair fireball becomes rapidly compressed into a shell of 
co-moving thickness $\Delta'$ with $R / \Delta' \sim \Gamma_0$ 
during the early evolution of the fireball, and that the 
ratio $R / \Delta'$ remains approximately constant after 
this compression phase, as long as the blast wave is coasting
(i.e. $\Gamma \approx$~const.). A relativistic pair fireball 
of this thickness, carrying a total energy per unit solid angle 
of $E_{\Omega} = 10^{52} \, E_{\Omega, 52}$~erg~sr$^{-1} = m_e 
c^2 \, \npair \, R^3 = 8.2 \times 10^{52} \, n'_{14} \, 
R_{15}^3$~erg~sr$^{-1}$ is initially highly optically thick
to Thomson scattering and thus radiatively inefficient since
radiation will remain trapped within the shell. Here, 
$\npair = 10^{14} \, n'_{14}$~cm$^{-3}$ is the co-moving
pair density (throughout the paper, primed quantities refer 
to the reference frame co-moving with the pair plasmoid), 
and $R_{15} = 10^{15} R_{15}$~cm, where $\sim 10^{15}$~cm 
is a typical value for the Thomson thinning radius (see 
\S \ref{results}). We assume that the magnetic field given by 
the equipartition parameter $\epsilon_B = 0.1 \, \epsilon_{B, -1}$ 
between magnetic field and pair plasma energy density (i.e., $u'_B = 
\epsilon_B \, u'_{\rm pair}$), and that the pair plasma is cold 
in the co-moving frame. Then, for a bulk Lorentz factor $\Gamma 
= 10^3 \, \Gamma_3$, the Alfv\'en crossing time scale in the 
observer's frame is

\begin{equation}
t_{\rm A} \sim 0.1 \, {R_{15} \over \Gamma_3^2 \, \epsilon_{B, -1}^{1/2}}
\; {\rm s}.
\label{t_A}
\end{equation}
Thus, Eq. (\ref{t_A}) indicates that it may take $\sim 
0.1$ -- 1~s, i. e. a non-negligible fraction of the prompt GRB 
phase, to build up the hydromagnetic turbulence for non-thermal 
particle acceleration at either internal or external shocks 
associated with these pair blastwaves. 

The time scale (in the observer's frame) for collisional processes, 
$t_{\rm coll}$, may be estimated as

\begin{equation}
t_{\rm coll} \sim \left( \Gamma \, c \, n'_{\rm pair} \, \sigma_T 
\right)^{-1} \sim 5 \times 10^{-4} \, \left( n'_{14} \, \Gamma_3
\right)^{-1} \; {\rm s},
\label{t_coll}
\end{equation}
indicating that during the initial phase --- prior to the generation
of strong hydromagnetic turbulence --- the radiative and energy-exchange
processes in the blast wave are dominated by collisional processes.
This fact is particularly interesting in view of the evidence in
a significant fraction of GRB that the early time-resolved
BATSE spectra show too hard a low-energy spectrum to be produced 
by non-thermal synchrotron emission \citep{crider97,preece98}.
Recently, \citet{preece01} has found evidence
in several BATSE GRBs that their very early spectra are described
better by thermal rather than nonthermal emission. 

Setting both time scales (\ref{t_A}) and (\ref{t_coll}) equal, one 
can estimate the condition under which non-thermal particle acceleration
may begin to dominate the particle energization behind the forward shock. 
Wwe find that collisional processes will dominate as long as

\begin{equation}
{t_A \over t_{\rm coll}} \sim 24 \, {E_{\Omega, 52} \over
\Gamma_3 \, R_{15}^2 \, \epsilon_{B, -1}^{1/2}} > 1.
\label{ta_tcoll}
\end{equation}

Although alternative emission models for the prompt GRB radiation have
been proposed by several authors (e.g., \cite{brainerd94,liang97}),
the dynamics and radiative signatures of the transition phase between
the time when the fireball becomes optically thin and the time of
efficient non-thermal particle acceleration by hydromagnetic turbulence
has not been studied in detail. In this paper, we are simulating the 
evolution of and time-dependent emission from a relativistic pair plasmoid
before the generation of hydromagnetic turbulence leading to non-thermal 
particle acceleration. In \S \ref{modelsetup}, we describe and motivate 
the basic assumptions of our model. The computation of the evolution of 
the thermal pair plasma and the fate of the swept-up background plasma 
are outlined in \S \ref{particleevolution}. In \S \ref{radiation} we 
describe the calculation of the time-dependent radiation spectra emitted 
by the resulting particle distributions in the blast wave. The evaluation
of the global dynamics of the blast wave is given in \S \ref{deceleration}.
Results of our numerical study are presented in \S \ref{results}. We
summarize in \S \ref{summary}.

\section{\label{modelsetup}Model setup}

We assume that the pair plasmoid forming the blastwave is a circular
shell segment with opening solid angle $\Omega$. Initially, it consists 
of a pure pair plasma with a pair density $n'_{\rm pair} = 10^{14} \, 
n'_{14}$~cm$^{-3}$ in the co-moving frame of the pair plasmoid. At the 
time when the pair fireball becomes optically thin to $\gamma\gamma$
pair production, the pairs are essentially cold in the co-moving frame. 
The blastwave interacts with an external medium (normal electron-proton 
plasma) of density $\nism = 1 \, \nismo$~cm$^{-3}$ in the stationary
frame of the surrounding medium. At any given time, the plasmoid is located 
at a radius $R = 10^{15} \, R_{15}$~cm from the center of the explosion 
and has a thickness $\Delta' = R/\Gamma$ in the co-moving frame.
The bulk Lorentz factor of the plasmoid is denoted by $\Gamma = 10^3 \, 
\Gamma_3$. The background magnetic field in the blastwave is 
$B = 1 \, B_0$~G and parametrized by the equipartition parameter
$\epsilon_B = \epsilon_{B, -1}$ through

\begin{equation}
B_0 = 1.4 \times 10^4 \, \left( \epsilon_{B, -1} n'_{14} \right)^{1/2}
\, \left( {K_3 \left[ {1 \over \Thpair} \right] \over K_2 \left[ {1 
\over \Thpair} \right]} - \Thpair \right)^{1/2}
\label{B0}
\end{equation}
where $\Thpair = kT_{\rm pair} / (m_e c^2)$ is the normalized 
(co-moving) pair temperature in the plasmoid, and $K_n$ is the
modified Bessel function of 2nd kind (a.k.a. McDonald function) 
of order $n$. For simplicity, we assume $\epsilon_B$ to be constant 
throughout our simulation. In fact, a gradual build-up of the 
magnetic field from lower initial values of $\epsilon_B$ may 
extend the period of applicability of our approach.

We are following the evolution of both the pair plasma as it is
being energized by Coulomb and inelastic (i.e. bremsstrahlung) 
collisions with swept-up electrons and protons from the surrounding
medium, and cools via radiative and adiabatic cooling. Simultaneously,
we solve for the evolution of the population of suprathermal protons
and electrons which are swept up from the environment with an initial 
Lorentz factor $\Gamma$ in the co-moving frame of the plasmoid, and 
transfer their energy to the background pair plasma and into radiation.

\section{\label{particleevolution}Evolution of pairs, electrons and 
protons in the blastwave}

Electrons and protons which are swept up by the pair blastwave have an
initial Lorentz factor $\Gamma$ in the co-moving frame, and will
transfer part of their kinetic energy to the background pairs in the 
plasmoid via elastic scattering (M\o ller/Bhabha and Coulomb scattering, 
respectively), thus heating the thermal pair plasma. The remainder of
the kinetic electron/proton energy will be radiated away through
bremsstrahlung with the background plasma and through synchrotron
radiation. The thermal background plasma will be heated by collisions
with the swept-up electrons and protons and cool via bremsstrahlung 
and synchrotron emission, adiabatic losses and pair annihilation. The
very short time scale for elastic scattering in the plasmoid (see
Eq. [\ref{t_coll}]) indicates that the incoming particles are rapidly
isotropized in the co-moving frame of the plasmoid. Thus, we assume
local isotropy of the particle distributions. 

\subsection{Suprathermal electron/proton cooling rates}

In the following, we quote simple approximation formulae for the cooling 
rates for suprathermal electrons and protons in a cold background pair 
plasma, which will be used in our simulations to follow the evolution
of the swept-up non-thermal particle distributions. For highly relativistic
test particles, the elastic scattering cross sections for electron-electron
(M\o ller) and electron-positron (Bhabha) scattering are approximately
equal, and we may use a simple power-law fit to the M\o ller scattering
energy exchange rate of a test particle with a cold background pair plasma 
given by \citet{nm98}. We find

\begin{equation}
\gdot_{\rm M\o, e} \approx 2.6 \times 10^3 \, n'_{14} \, \gamma_e^{-0.02} \, 
{\rm s}^{-1}.
\label{dge_mo}
\end{equation}

The \ee and \ep bremsstrahlung energy loss rate can be approximated as 

\begin{equation}
\gdot_{\rm br, e} \approx 8 \times 10^{-2} \, n'_{14} \, \gamma_e^{1.15} \, 
{\rm s}^{-1}.
\label{dge_br}
\end{equation}
Suprathermal electrons will also emit synchrotron radiation, thus
losing energy at a rate

\begin{equation}
\gdot_{\rm sy, e} \approx 1.29 \times 10^{-9} \, B_0^2 \, \gamma_e^2 \, 
{\rm s}^{-1}.
\label{dge_sy}
\end{equation}
Finally, we take into account synchrotron-self-Compton emission and
estimate the associated energy loss as

\begin{equation}
\gdot_{\rm SSC, e} \approx - {4 \over 3} \, c \, \sigma_{\rm T} \,
{u'_{\rm sy, e} \over m_e c^2} \, \gamma_e^2
\label{dge_ssc}
\end{equation}
where $u'_{\rm sy} \approx L'_{\rm sy, e} / (\Omega \, R^2 \, c)$ is the
energy density in synchrotron photons. These estimates indicate that
except for the highest-energy electrons (with $\gamma_e \gs 10^3$),
the energy loss will be dominated by elastic scattering, leading to
rapid thermalization of the suprathermal electrons.

For Coulomb losses of suprathermal protons in the cold pair plasma, we 
use Eq. (4.22) of \citet{ms94} with
$\beta \gg \beta_e$ (i.e. ultrarelativistic protons in a non-relativistic 
or mildly relativistic background plasma): 
\begin{equation}
\gdot_{\rm Coul, p} \approx 3.3 \times 10^{-2} \, n'_{14} \, {\rm s}^{-1}
\label{dgp_Coul}
\end{equation}

The bremsstrahlung energy loss rate of suprathermal protons is evaluated
by integrating the respective photon spectrum calculated by citet{jones71} 
over the outgoing photon energies. Since to our knowledge, the result has 
not yet been published elsewhere, we derive the complete expression in 
Appendix \ref{pbrloss}. In our simulations, we use a simple power-law 
fit to the resulting bremsstrahlung energy loss rate:
\begin{equation}
\gdot_{\rm br, p} \approx 1.1 \times 10^{-5} \, n'_{14} \, \gamma_p^{0.3} \,
{\rm s}^{-1}
\label{dgp_br}
\end{equation}
Finally, we take into account synchrotron emission of ultrarelativistic
protons, leading to an energy loss of

\begin{equation}
\gdot_{\rm sy, p} \approx {c \, \sigma_T \, B^2 \over 6 \pi \, m_p c^2}
\, \left( {m_e \over m_p} \right)^2 \, \gamma^2 \approx 2.09 \times 10^{-19}
\, B_0^2 \, \gamma_p^2 \, {\rm s}^{-1}.
\label{dgp_sy}
\end{equation}
As in the case of suprathermal electrons, we expect that the swept-up 
protons will lose energy primarily via elastic scattering with the
thermal background pair plasma.

\subsection{Suprathermal electron and proton spectra}

As mentioned in the previous section, we generally expect that the 
energy loss of both protons and electrons might be dominated by 
Coulomb interactions for $\gamma_{\rm p, max} = \Gamma \ls 10^3$. 
However, for the highest-energy protons and electrons, synchrotron 
losses may also play an important role. The evolution of the suprathermal
electron and proton populations may be calculated by solving the continuity 
equation for the cooling particles: 

\begin{equation}
{\partial N_e (\gamma, t') \over \partial t'} = - {\partial \over \partial
\gamma} \left( \dot\gamma \, N[\gamma, t'] \right) + \dot N_{\rm sw} (\gamma,
t'),
\label{continuity}
\end{equation}
where $\dot N_{\rm sw} (\gamma, t') = \beta_{\Gamma} c \, \Gamma 
\, n_{\rm ISM} \Omega \, R^2 \, \delta(\gamma - \Gamma)$ is the 
sweep-up rate of ISM particles, and we have neglected escape. Since 
the cooling time scale for ultrarelativistic particles is of the 
order of or shorter than the dynamical time scale, we may approximate 
the evolution by a sequence of quasi-equilibrium solutions to 
Eq. (\ref{continuity}), given by

\begin{equation}
N_e (\gamma_e) = N_e^0 \cases{
\left( {\gamma_e \over \gamma_e^{\rm br}} \right)^{0.02} & 
for $1 \le \gamma_e \le \gamma_e^{\rm br}$ \cr\cr
\left( {\gamma_e \over \gamma_e^{\rm br}} \right)^{-2} & 
for $\gamma_e^{\rm br} \le \gamma_e \le \Gamma$ \cr}
\label{Ne}
\end{equation}
for the electrons and

\begin{equation}
N_p (\gamma_p) = N_p^0 \cases{ 1 & 
for $1 \le \gamma_p \le \gamma_p^{\rm br}$ \cr\cr
\left( {\gamma_p \over \gamma_p^{\rm br}} \right)^{-2} & 
for $\gamma_p^{\rm br} \le \gamma_p \le \Gamma$ \cr}
\label{Np}
\end{equation}
for the protons. The break energies are defined through the condition
$\gdot_{\rm elast. \, scat.} = \gdot_{\rm sy}$ and given by

\begin{equation}
\gamma_e^{\rm br} = 8.5 \times 10^6 \, (n'_{14})^{1 \over 2.02} \, 
B_0^{-{1 \over 1.01}}
\label{gbr_e}
\end{equation}
and
\begin{equation}
\gamma_p^{\rm br} = 3.95 \times 10^8 \, (n'_{14})^{1/2} \, B_0^{-1}.
\label{gbr_p}
\end{equation}

If $\gamma_{e/p}^{\rm br} \ge \Gamma$, the high-energy branches of
Eqs. (\ref{Ne}) and (\ref{Np}) are not realized, and the respective
particle spectra are single truncated power-laws for $1 \le \gamma_{e/p}
\le \Gamma$.

The background pair plasma is assumed to maintain a thermal distribution
at a temperature determined by the balance between heating by the
swept-up particles and radiative cooling via bremsstrahlung and synchrotron
emission.

\subsection{Temperature of thermal pair plasma}

As mentioned above, most of the kinetic energy of the protons is
transferred to electrons through Coulomb collisions with the thermal
background pair plasma. Thus, the heating rate of the thermal plasma 
is given by

\begin{equation}
L_{\rm heat} = \Omega \, R^2 \, \Gamma^2 \, m_p c^2 \, \beta \, c \, \nism
\approx 4.50 \times 10^{41} \, \Omega \, R_{15}^2 \, \Gamma_3^2 \, \nismo
\; {\rm erg \; s}^{-1}.
\label{L_heat}
\end{equation}

This heating should be balanced by cooling through bremsstrahlung,

\begin{equation}
L_{\rm br, th} \approx \cases{ 3.42 \times 10^{48} \, \Omega \, (n'_{14})^2
\, R_{15}^2 \, \Delta'_{12} \, \sqrt{\Thpair} \; {\rm erg \; s}^{-1} &
for $\Thpair \ll 1$ \cr\cr
1.36 \times 10^{49} \, \Omega \, (n'_{14})^2 \, R_{15}^2 \, \Delta'_{12} \,
\Thpair \left( \ln[2 \Thpair] + 0.673 \right) \; {\rm erg \; s}^{-1} &
for $\Thpair \gs 1$ \cr}
\label{L_br_th}
\end{equation}
--- where $\Delta'_{12} = \Delta' / (10^{12} \; {\rm cm})$ ---,
synchrotron emission,

\begin{equation}
L_{\rm sy, th} \approx 3.175 \times 10^{41} \, \Omega \, n'_{14} \, R_{15}^2 
\, \Delta'_{12} \, B_0^2 \, \Theta \, {K_3 (1/\Theta) \over K_2 (1/\Theta)}
\; {\rm erg \; s}^{-1}
\label{L_sy_th}
\end{equation}
(cf. \cite{bps99}), and thermal Comptonization of the soft photon field
(at $E \ll m_e c^2$). For typical parameters, the pair plasmoid is expected
to be moderately optically thick, $\tau_{\rm T} \gs 1$, and have
mildly relativistic temperatures, $\Theta \ls 1$. In this regime,
we may use the analytic solution of \citet{ht95}
for the saturated-Comptonization case (their Eq. [9]). We assume that the
dominant soft input radiation field is the electron synchrotron component,
which we approximate, for simplicity, by a $\delta$ function spectrum \,
$L_{\rm s} (\epsilon) = L_{\rm sy, e} \, \delta (\epsilon - \epsilon_{\rm
sy, e}^{\rm max})$. The cooling rate due to Thermal Comptonization is
then calculated by integrating the Comptonized spectrum (in the comoving
frame) over photon energy.

Adiabatic cooling is described by a virtual luminosity $L_{\rm adi} 
= 3 \, E_{\rm th} \, c / R$, where $E_{\rm th}$ is the thermal
energy content of the thermal pair plasma. At nonrelativistic pair 
plasma temperatures, pair annihilation becomes important. We use the 
expressions of \citet{svensson82} for the pair 
annihilation cooling rate and the pair annihilation rate to calculate 
the cooling and depletion of the number of pairs due to this process.

\section{\label{radiation}Radiation spectra}

The radiation emitted by the thermal pair plasma can be calculated using
standard expressions (see, e.g., \cite{bps99}, and \cite{ht95} for the
Thermal Comptonization component). The intrinsic luminosities of the 
bremsstrahlung and synchroton components are given by Eqs. (\ref{L_br_th}) 
and (\ref{L_sy_th}), respectively.

The (optically thin) bremsstrahlung spectrum of the suprathermal electrons 
in the co-moving frame reproduces the broken power-law shape of the 
electron spectrum:

\begin{equation}
L'_{\rm br, e} (\epsilon') = L_{\rm br, e}^0 \, \cases{
1 & for $\epsilon' \le \epsilon_{\rm br, e}^{\rm br}$ \cr\cr
\left( {\epsilon' \over \epsilon_{\rm br, e}^{\rm br}} \right)^{-1} & for
$\epsilon_{\rm br, e}^{\rm br} \le \epsilon' \le \Gamma$ \cr}
\label{e_br_spectrum}
\end{equation}
where $\epsilon_{\rm br, e}^{\rm br} = \gamma_e^{\rm br}$ and $\epsilon' =
h \nu' / (m_e c^2)$. The suprathermal proton bremsstrahlung 
spectrum may be approximated as

\begin{equation}
L'_{\rm br, p} (\epsilon) = L_{\rm br, p}^0 \, \cases{
1 & for $\epsilon' \le {1 \over 2}$ \cr\cr
(2 \, \epsilon')^{-1.25} & for
${1 \over 2} \le \epsilon' \le \Gamma$ \cr}
\label{p_br_spectrum}
\end{equation}
\citep{jones71}.

The nonthermal synchrotron spectra can be approximated as

\begin{equation}
L'_{\rm sy, e/p} (\epsilon') = L_{\rm sy, e/p}^0 \, \cases{
\left( {\epsilon' \over \epsilon_{\rm sy, e/p}^{\rm br}} \right)^{1/3}
& for $\epsilon' \le \epsilon_{\rm sy, e/p}^{\rm br}$ \cr\cr
\left( {\epsilon' \over \epsilon_{\rm sy, e/p}^{\rm br}} \right)^{-1/2}
& for $\epsilon_{\rm sy, e/p}^{\rm br} \le \epsilon' \le 
\epsilon_{\rm sy, e/p}^{\rm max}$
\cr}
\label{nth_sy_spectrum}
\end{equation}
where 

\begin{equation}
\epsilon_{\rm sy, e}^{\rm br} = {B \over B_{\rm c}} \, (\gamma_e^{\rm br})^2
\label{eebr}
\end{equation}
\begin{equation}
\epsilon_{\rm sy, e}^{\rm max} = {B \over B_{\rm c}} \, \Gamma^2
\label{eemax}
\end{equation}
\begin{equation}
\epsilon_{\rm sy, p}^{\rm br} = {3 \over 2} \, {m_e \over m_p} \, {B \over 
B_{\rm c}} ( \gamma_p^{\rm br} )^2
\label{epbr}
\end{equation}
\begin{equation}
\epsilon_{\rm sy, p}^{\rm max} = {3 \over 2} \, {m_e \over m_p} \, {B \over 
B_{\rm c}} \Gamma^2
\label{epmax}
\end{equation}
with $B_{\rm c} = 4.414 \times 10^{13}$~G. 

Using a simple delta-function approximation for the Compton-scattering
cross section in the Thomson limit, the evaluation of the electron-SSC 
spectrum is straightforward, but a bit lengthy. The full expression is
given in Appendix \ref{sscspectrum}. 

The normalization factors of the radiation spectra are given by setting the 
energy-integrated luminosities,

\begin{equation}
L_{\rm br, e} = L_{\rm br, e}^0 \, \epsilon_{\rm br, e}^{\rm br} \left(
1 + \ln \left[ {\Gamma \over \epsilon_{\rm br, e}^{\rm br}} \right] \right)
\label{Lbrenorm}
\end{equation}
\begin{equation}
L_{\rm br, p} = {L_{\rm br, p}^0 \over 2} \, \left(
1 + 4 \left[1 - (2 \Gamma)^{-0.25} \right] \right)
\label{Lbrpnorm}
\end{equation}
\begin{equation}
L_{\rm sy, e/p} = L_{\rm sy, e/p}^0 \, \left({4 \over 3} \, 
\epsilon_{\rm sy, e/p}^{\rm br} + 2 \, \sqrt{\epsilon_{\rm sy, e/p}^{\rm br}}
\left[ \sqrt{ \epsilon_{\rm sy, e/p}^{\rm max}} - \sqrt{ \epsilon_{\rm
sy, e/p}^{\rm br}} \right] \right)
\label{Lsynorm}
\end{equation}
\begin{equation}
L_{\rm SSC, e} = L_{\rm SSC}^0 \, N 
\label{Lsscn}
\end{equation}
where $N$ is given in Appendix \ref{sscspectrum} (Eq. [\ref{N_SSC}]), 
equal to the fraction of swept-up kinetic particle energy, which 
is transferred into the respective radiative cooling channel,

\begin{equation}
L_{\rm br, e} = \Omega \, \Gamma^2 \, R^2 \, m_e c^2 \, \beta \, c \, 
\nism \, {\gdot_{\rm br, e} \over \gdot_{\rm M\o, e} + \gdot_{\rm br, e} 
+ \gdot_{\rm sy, e} + \gdot_{\rm SSC, e}}
\label{Lbre}
\end{equation}
\begin{equation}
L_{\rm br, p} = \Omega \, \Gamma^2 \, R^2 \, m_p c^2 \, \beta \, c \, 
\nism \, {\gdot_{\rm br, p} \over \gdot_{\rm Coul, p} + \gdot_{\rm br, p} 
+ \gdot_{\rm sy, p}}
\label{Lbrp}
\end{equation}
\begin{equation}
L_{\rm sy, e} = \Omega \, \Gamma^2 \, R^2 \, m_e c^2 \, \beta \, c \, 
\nism \, {\gdot_{\rm sy, e} \over \gdot_{\rm M\o, e} + \gdot_{\rm br, e} 
+ \gdot_{\rm sy, e} + \gdot_{\rm SSC, e}}
\label{Lsye}
\end{equation}
\begin{equation}
L_{\rm sy, p} = \Omega \, \Gamma^2 \, R^2 \, m_p c^2 \, \beta \, c \, 
\nism \, {\gdot_{\rm sy, p} \over \gdot_{\rm Coul, p} + \gdot_{\rm br, p} 
+ \gdot_{\rm sy, p}}
\label{Lsyp}
\end{equation}
\begin{equation}
L_{\rm SSC, e} = \Omega \, \Gamma^2 \, R^2 \, m_e c^2 \, \beta \, c \, 
\nism \, {\gdot_{\rm SSC, e} \over \gdot_{\rm M\o, e} + \gdot_{\rm br, e} 
+ \gdot_{\rm sy, e} + \gdot_{\rm SSC, e}}
\label{Lsssce}
\end{equation}
All luminosities have been calculated in the co-moving frame 
and are transformed to apparent isotropic luminosities through 

\begin{equation}
L_{\rm app} (\epsilon) = f_L \, {1 - \exp^{- \tau_{\gamma\gamma}
(\epsilon')} \over \tau_{\gamma\gamma} (\epsilon')} \, D^3 \, 
L' (\epsilon'),
\label{L_boost}
\end{equation}
where $D = \left( \Gamma [1 - \beta_{\Gamma} 
\cos\theta_{\rm obs}] \right)^{-1}$ is the Doppler beaming 
factor, $\epsilon = D \, \epsilon'$, and $f_L = \min (1, \, 
c t' / \Delta') \times \min( 1, \, 1/[\Gamma^2 \Omega])$ is 
a correction factor accounting for the following two effects: 
First, due to light-travel-time constraints, only emission from 
a fraction of $\sim c t' / \Delta'$ of the plasmoid will be 
visible to the observer. Second, if the opening angle of the 
cone in which plasmoid is moving, is larger than the beaming 
angle $1 / \Gamma$, then only a fraction $1 / (\Gamma^2 \, \Omega)$ 
of the surface will contribute to the observed emission. In this
analysis, we neglect the fact that in reality the observed emission
is a superposition of emission from different layers of the plasmoid
which have different evolutionary ages. This effect may alter the
detailed spectral shape and light curves slightly (see, e.g., 
\citet{pm98} for a detailed
analysis of this effect in the case of GRB afterglows), but will 
not change the qualitative results of our analysis.

The effect of pair production by $\gamma\gamma$ absorption is
treated by injecting a number of thermal pairs equal to the
number of absorbed photon pairs within any given time step 
into the plasmoid. Since we have shown that thermalization is
the dominant energy-loss process for suprathermal particles,
this approach yields an accurate description of the 
$\gamma\gamma$ pair production process in the plasmoid.

\section{\label{deceleration}Plasmoid deceleration}

Due to momentum conservation, the bulk Lorentz factor $\Gamma$ of the
pair plasmoid will decrease as it sweeps up external matter, according
to

\begin{equation}
d\Gamma = - {(\Gamma^2 - 1) \, dm + \Gamma \, dE_{\rm adi} \over M}
\label{dGamma}
\end{equation}
\citep{dh01} where $dm = \Omega \, R^2 \, (m_p + m_e) \, n_{\rm ISM} 
\, \beta \, c \; dt$, $dE_{\rm adi}$ is the loss of internal energy
due to adiabatic cooling, and $M$ is the total, relativistic mass in 
the fireball (i. e. rest mass + internal kinetic energy of pairs + 
mass of swept-up external material). The rest-mass increment $dm$ is 
related to the increment in $M$ by

\begin{equation}
dM = \Gamma \, dm - {dE_{\rm rad} + dE_{\rm adi} \over c^2}
\label{dM}
\end{equation}
where $dE_{\rm rad}$ is the net energy produced in radiation throughout
the plasmoid.

\section{\label{results}Numerical results}

For a general parameter study, we have performed a series of pair 
plasmoid simulations with various values of bulk Lorentz factors
$\Gamma$, co-moving pair densities $\npair$, initial radii $R_0$,
and external matter densities $\nism$. Throughout our simulations, 
we have fixed the magnetic field equipartition parameter to 
$\epsilon_B = 0.1$, and the observer is assumed to be located 
on the symmetry axis, i.e. $\theta_{\rm obs} = 0$. We keep the 
ratio $R / \Delta' = \Gamma_0$ constant at its initial value. 

\begin{figure}
\plotone{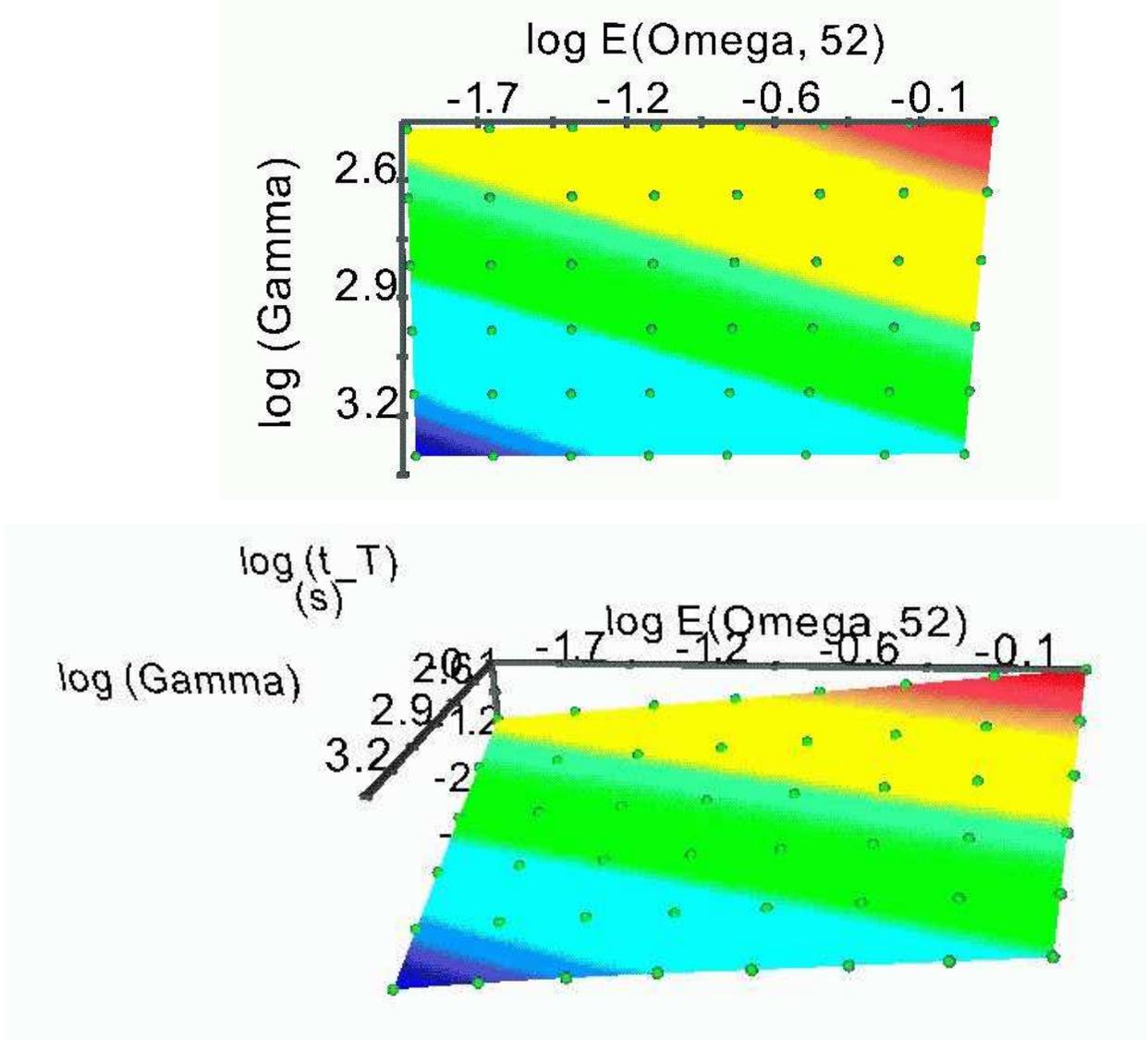}
\caption{Thomson thinning time $t_{\rm T}$ of a relativistic pair plasmoid
injected at $R_0 = 10^{14}$~cm, as a function of total injected energy per
unit solid angle, $E_{\Omega}$, and initial bulk Lorentz factor $\Gamma_0$. 
Other parameters: $n_{\rm ISM} = 100$, and $R / \Delta' = \Gamma$. The 
green dots indicate simulated values; the surface has been constructed 
using a spline interpolation. The surface colors encode the values of
$t_T$ (vertical axis in the lower panel) with red corresponding to $t_T 
\sim 0.5$ -- 1~s and blue corresponding to $t_T \sim 0.5$ -- 1~ms.}
\label{tT_figure}
\end{figure}

As briefly mentioned in the introduction, our simulations generally
start out in a highly Thomson thick regime, in which the output is
basically a thermal blackbody at the blue-shifted pair temperature.
For this reason, the results are virtually independent of the initial
radius $R_0$, and for the results presented in the following, we 
have chosen $R_0 = 10^{14}$~cm. In this initial phase, the 
plasmoid is inefficient in terms of high-energy emission. 
At the radius where the Thomson depth is $\sim 1$, the plasmoid 
becomes radiatively efficient, and a flash of high-energy 
$\gamma$-ray emission is produced. We denote the time at which the
plasmoid becomes marginally Thomson-thin as $t_{\rm T}$. This
transition time is plotted as a function of the essential model
parameters $E_{\Omega, 52}$ and $\Gamma_0$ in Fig. \ref{tT_figure}
which shows that it gradually increases with increasing injected
pair energy and decreasing bulk Lorentz factor. We find typical
values of $0.01$~s~$\ls t_T \ls 0.1$~s. 

\begin{figure}
\plotone{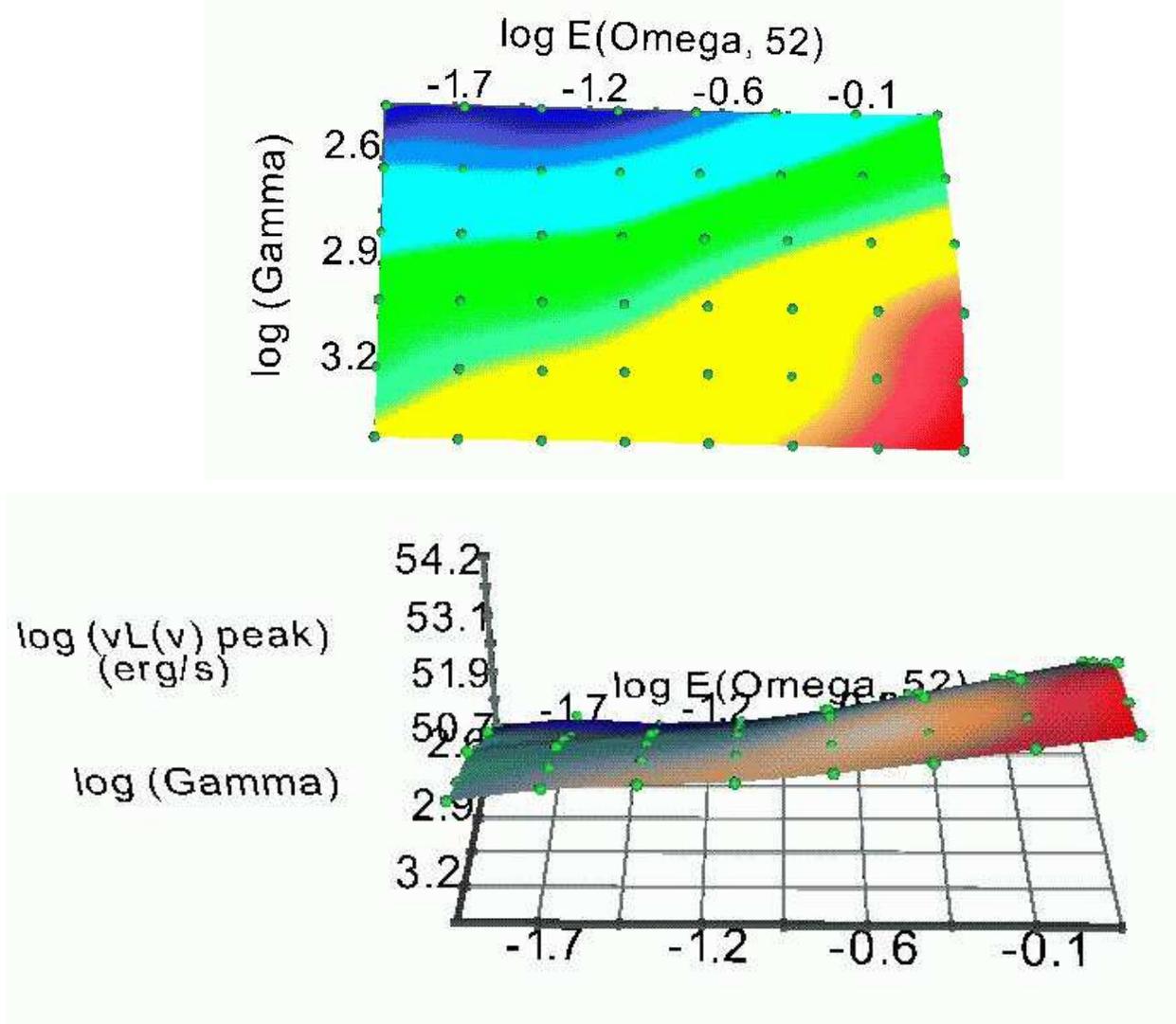}
\caption{Peak $\nu L_{\nu}$ apparent isotropic luminosity at the time 
of maximum received flux, as a function of $E_{\Omega, 52}$ and 
$\Gamma_0$. Parameters are the same as in Fig. \ref{tT_figure}. 
The green dots indicate simulated values. The surface colors encode 
the values of $\nu L_{\nu, {\rm peak}}$ with red corresponding to 
$\nu L_{\nu, {\rm peak}} \sim 10^{54}$~ergs~s$^{-1}$ and blue 
corresponding to $\nu L_{\nu, {\rm peak}} \sim 10^{51}$~ergs~s$^{-1}$.}
\label{nuLnu_figure}
\end{figure}

\begin{figure}
\plotone{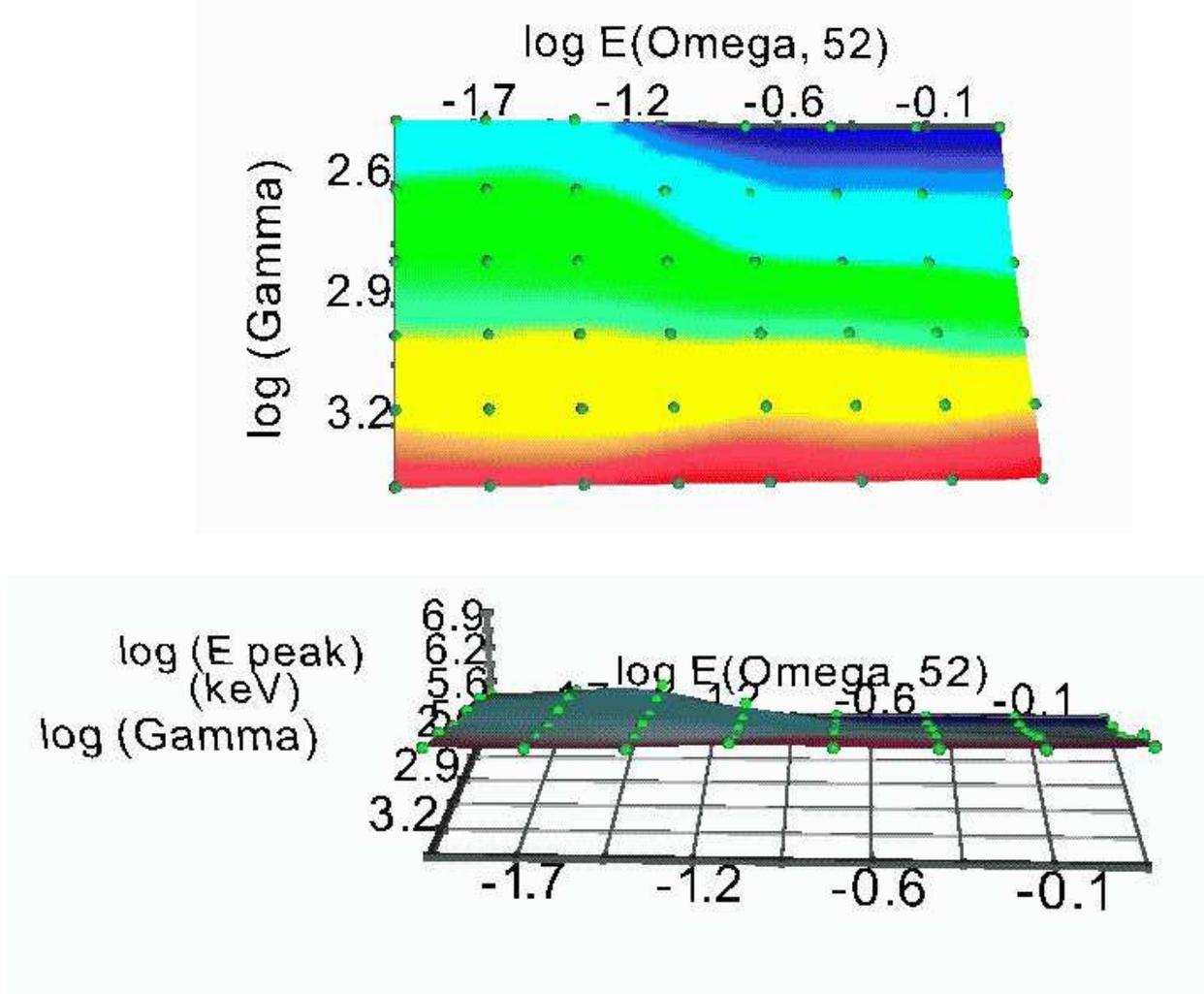}
\caption{Peak photon energy at the time of maximum received
flux, as a function of $E_{\Omega, 52}$ and $\Gamma_0$. 
Parameters are the same as in Fig. \ref{tT_figure}. The 
green dots indicate simulated values. The surface colors encode 
the values of $E_{\rm peak}$ with red corresponding to $E_{\rm peak} 
\sim 10$~GeV and blue corresponding to $E_{\rm peak} \sim 100$~MeV.}
\label{Epk_figure}
\end{figure}

For each simulation, we then evaluate, among other quantities, the 
apparent quasi-isotropic $\nu L_{\nu}$ peak luminosity and the peak 
photon energy $E_{\rm pk}$ at the time of maximum received flux. The
dependence of those quantities on the parameters $E_{\Omega, 52}$ 
and $\Gamma_0$ is illustrated in Figs. \ref{nuLnu_figure} and 
\ref{Epk_figure}. 

In these calculations, we have fixed $n_{\rm ISM} = 100$~cm$^{-3}$. 
The graphs show that we find typical peak $\nu L_{\nu}$ luminosities
$\sim 10^{52}$ -- $10^{54}$~ergs~s$^{-1}$, increasing with increasing
$E_{\Omega, 52}$ and $\Gamma$. $\nu F_{\nu}$ peak energies $E_{\rm pk}$
are typically in the range $\sim 100$~MeV -- several~GeV, and increase
with $\Gamma$, while they are only weakly dependent on the injected
energy $E_{\Omega, 52}$.

\begin{figure}
\plotone{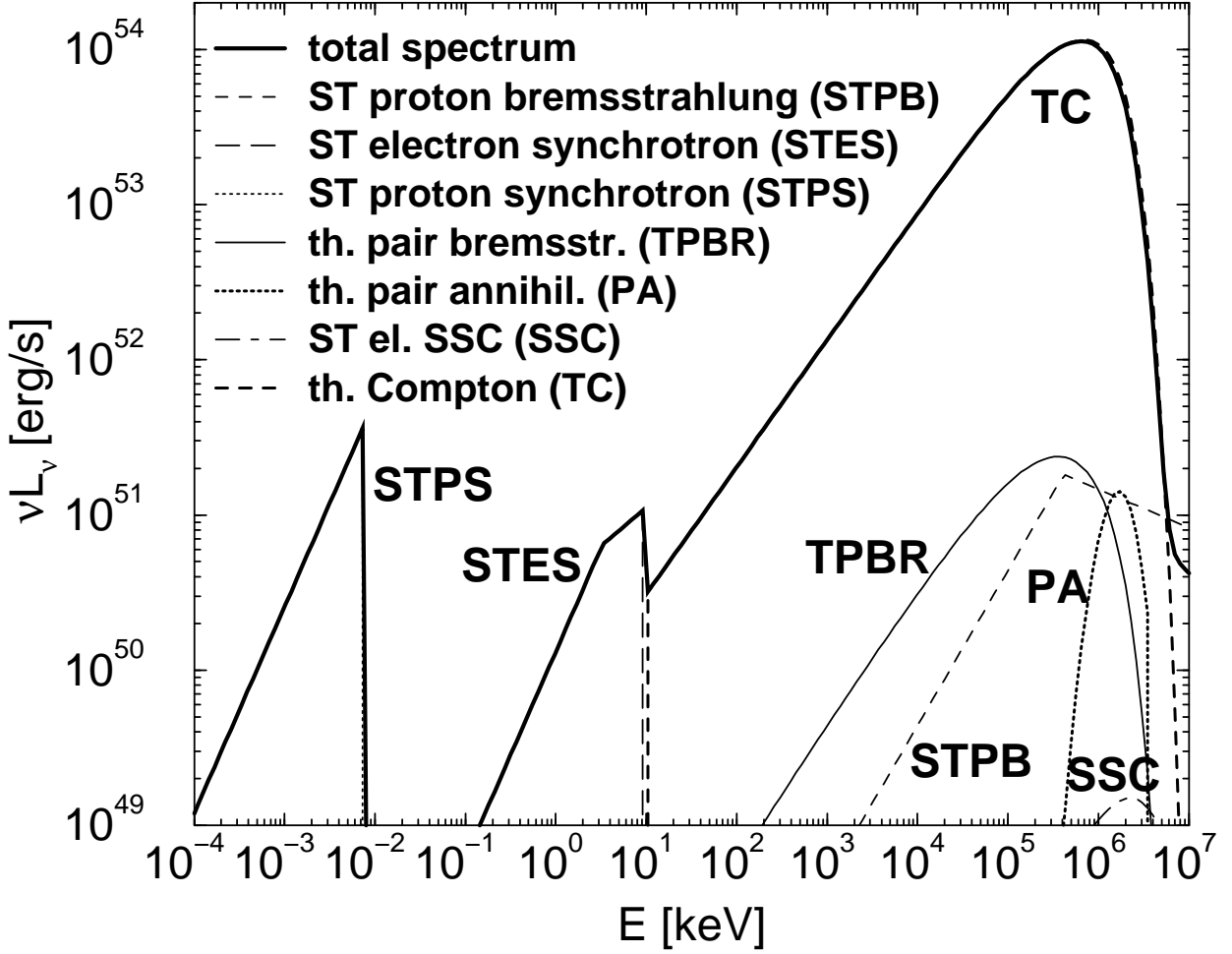}
\caption{Composite photon spectrum from the blast wave at the time of 
maximum received flux ($t_{\rm rec} = 64$~ms). Parameters for this
simulation are: $\Gamma_0 = 10^3$, $\npair = 1.2 \times 10^{16}$~cm$^{-3}$, 
$R_0 = 10^{14}$~cm, implying $E_{\Omega, 52} = 1$; $n_{\rm ISM} = 
100$~cm$^{-3}$, $\epsilon_B = 0.1$, $\theta_{\rm obs} = 0$. $\gamma\gamma$ 
absorption has been taken into account to calculate the total emission, 
while the individual contributions are plotted without correction for 
$\gamma\gamma$ absorption. Thus, the importance of $\gamma\gamma$ 
absorption and pair production is illustrated by the difference between 
the individual, unabsorbed components and the total, emerging spectrum 
at $E \gs 1$~GeV.}
\label{composite2}
\end{figure}

For a typical simulation in this range of parameters, Fig. \ref{composite2}
shows the composite spectrum with the individual radiation components
at the time of maximum observed flux. The X-ray and $\gamma$-ray spectrum
is clearly dominated by thermal Comptonization in this phase. 
time corresponding to Fig. \ref{composite2}, the radial Thomson depth 
of the plasmoid is $\tau_{\rm T} \sim 0.24$; the pair temperature
is $\Thpair \sim 0.44$. The spiky shape of the electron and proton 
synchrotron spectra is due to the $\delta$ function approximation 
to the respective emissivities. 

\begin{figure}
\plotone{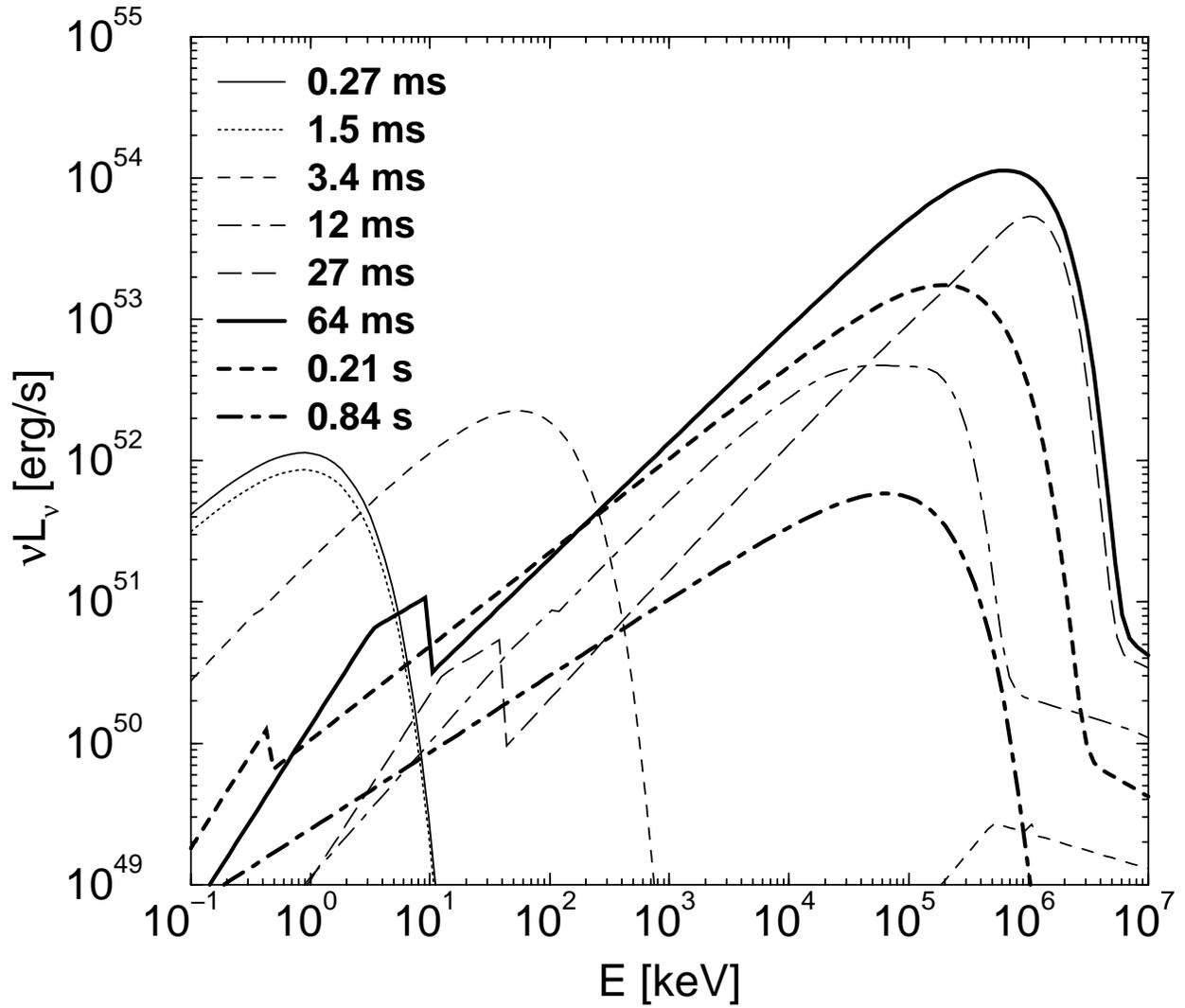}
\caption{Time evolution of the observed photon spectra for the same
set of parameters as in Fig. \ref{composite2}.}
\label{evol2}
\end{figure}

Fig. \ref{evol2} illustrates the spectral evolution during the
collision dominated phase of the pair plasmoid evolution. 
It shows that the hard X-ray and $\gamma$-ray spectral output 
remains dominated by thermal Comptonization throughout most of 
this phase. During the first few ms, there is a strong soft 
X-ray component due to the thermal emission of the cold pair
plasmoid, which is rapidly shifting to higher energies as the
pairs are being heated. At very high energies, $E \gs 1$~GeV, 
there is a weak tail from suprathermal proton bremsstrahlung. 
According to Eq. (\ref{t_A}), hydromagnetic turbulence might 
develop within $\sim 1$~s in the case discussed here. 

\begin{figure}
\plotone{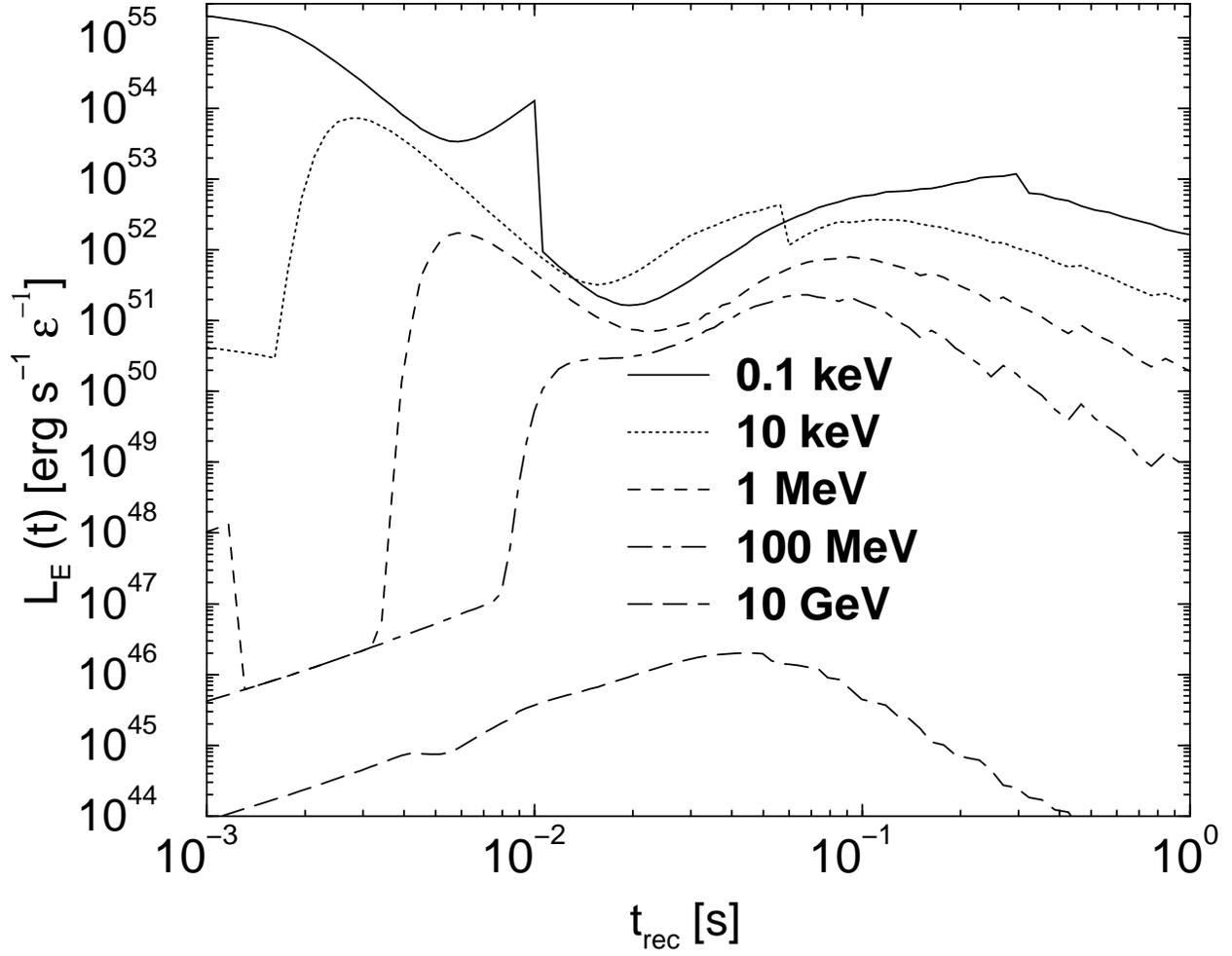}
\caption{Light curves at several different observed photon energies
for the same set of parameters as in Fig. \ref{composite2}.}
\label{lc2}
\end{figure}

Light curves expected from the collision-dominated pair plasmoid
are shown in Fig. \ref{lc2}. The figure illustrates that the
maximum spectral power output is expected around $\sim 0.1$~s
after the formation of the pair plasmoid. Generally, the time
of maximum flux decreases with increasing photon energy,
indicating an overall hard-to-soft spectral evolution. 

\section{\label{summary}Summary and conclusions}

We have investigated the early stages of the evolution of a 
relativistic pair fireball, immediately after it has become 
optically thin to $\gamma\gamma$ pair production. We have shown 
that for a short period of time, $\sim 0.1$ -- a few seconds 
after the initial explosion, the pair blast wave evolution might 
be dominated by collisional processes prior to the formation of a 
collisionless shock which would subsequently lead to non-thermal
particle acceleration at the shock front. We have simulated the 
relevant energy exchange and radiation processes, including
suprathermal electron and proton thermalization and bremsstrahlung,
suprathermal electron and proton synchrotron emission, thermal
pair bremsstrahlung, thermal pair annihilation, thermal Comptonization,
and $\gamma\gamma$ pair production and absorption, during the early 
pair plasmoid evolution, and calculated the expected radiative 
signatures. 

We have investigated the dependence of the radiative signatures on
pair plasmoid parameters which are very hard to predict from first
principles in the framework of current GRB progenitor models. 
In the range of total injected energies of $E_{\Omega} \sim 
10^{52}$~ergs~sr$^{-1}$ and initial bulk Lorentz factors $\Gamma_0 
\sim 500$ -- 1000, the radiation from the collision-dominated 
pair plasmoid phase results in a short period ($\sim$~a few ms) 
of thermal soft X-ray emission from the initially Thomson thick, 
cold pair plasmoid. As the pair plasma is heated due to the
sweeping-up of external material and is becoming Thomson thin
due to expansion, the observable emission turns into a quasi-thermal 
spectrum, peaking at around $E_{\rm pk} \sim$~100 MeV -- a few 
GeV, dominated by thermal Comptonization by the mildly relativistic 
pair plasma in the plasmoid during the first $\sim 0.01$ -- 1~sec 
after the onset of the GRB. The apparent peak $\nu L_{\nu}$ 
luminosities are expected in the range of $\sim 10^{52}$ -- 
$10^{54}$~ergs/s, sustained over typically a few tens of 
milliseconds. 

The expected very early, thermal signatures may already have been 
observed in time-resolved, early BATSE spectra of a few GRBs. The
quasi-thermal spectra expected during the optical-thinning transition
phase, $\sim 0.01$ -- 1~s after the onset of the GRB, are peaking 
in the $\sim 100$~MeV regime. Thus, they may have remained undetectable
for BATSE, but should be easily detectable by the GLAST mission, 
scheduled for launch in 2005.

\acknowledgments
We thank the anonymous referee for helpful and constructive
comments, and Dr. C. D. Dermer for careful reading of the 
manuscript and stimulating discussions. The work of MB is 
supported by NASA through Chandra Postdoctoral Fellowship 
grant PF~9-10007 awarded by the Chandra X-ray Center, which 
is operated by the Smithsonian Astrophysical Observatory 
for NASA under contract NAS~8-39073. RS acknowledges partial
support by the Bundesministerium f\"ur Bildung und Forschung
through DESY, grant 05AG9PCA.

\appendix

\section{\label{pbrloss}Bremsstrahlung losses of suprathermal protons}

In order to calculate the energy loss rate of suprathermel electrons
due to bremsstrahlung in inelastic collisions with a cold pair
plasma, we start out with the differential cross section $d\sigma
\over d\epsilon$ as given in Eq. (4) of \citet{jones71}
(for consistency we use $\epsilon$ for the normalized energy of the
outgoing photon, which is $\alpha$ in Jones' notation) and evaluate

\begin{equation}
\gdot_{\rm br, p} = - n'_e \, \beta_p \, c \, {m_e \over m_p} 
\int\limits_0^{\epsilon_{\rm max}} d\epsilon \; \epsilon \, {d\sigma 
\over d\epsilon}
\label{gbrp}
\end{equation}
where
\begin{equation}
\epsilon_{\rm max} = 1.123 \, \gamma_p \beta_p^2 \, e^{-{\beta_p^2 \over 2}}.
\label{A}
\end{equation}
Evaluating the integral in (\ref{gbrp}), we find

\begin{equation}
\gdot_{\rm br, p} = - {3 \over 4} {\sigma_T \, \alpha \, c \over \pi}
{m_e \over m_p} \, {n'_e \over \beta_p} \, G(\gamma_p)
\label{gdotpbr}
\end{equation}
where $\alpha \approx 1/137$ is the fine structure constant, and
\begin{equation}
G(\gamma_p) = {2 \over 3} \left( \ln A \right)^2 + \left( {4 \over 3}
\ln 2 + {13 \over 18} \right) \, \ln A + {190 \over 27} - {2 \over 9}
\pi^2 + {19 \over 40} \ln 2 + O \left( {1 \over A} \right),
\label{G}
\end{equation}
where $A \approx 0.68 \gamma_p \gg 1$ for relativistic protons. The 
last term in Eq. (\ref{G}) denotes all terms of order $1 / A \ll 1$.

\section{\label{sscspectrum}The electron-SSC spectrum}

We evaluate the electron-SSC spectrum using a $\delta$ function 
approximation for the Compton scattering cross section in the 
Thomson regime \citep{reynolds82}, with a sharp cut-off at
incident photon energies $\epsilon\gamma_e \ge 1$. In the case 
$\gamma^{\rm br}_e = \Gamma$ this yields:

\begin{equation}
L_{\rm SSC} (\epsilon) = L_{\rm SSC}^0 \, \epsilon^{1/3} \, \left(
\Gamma^{1/3} - \Gamma^{-1/3} \left[ {3 \over 4} \epsilon {B_{\rm cr}
\over B} \right]^{1/6} \right) \> \Theta\left( {4 \over 3} \Gamma^4 \,
{B \over B_{\rm cr}} - \epsilon \right)
\label{ssc_sp_I}
\end{equation}
In the case $\gamma^{\rm br}_e < \Gamma$, we find:

\begin{equation}
L_{\rm SSC} (\epsilon) = L_{\rm SSC}^0 \cdot \cases{
3 \, \left({4 \over 3}\right)^{2/3} \, \left({\epsilon \over 
\epsilon_{\rm sy, e}^{\rm br}}\right)^{1/3} \, (\gamma_e^{\rm br})^{1/3} & \cr
\cdot \, \left(1 + {1 \over 5} \left[ 1 - \left\lbrace {\gamma_e^{\rm br} 
\over \Gamma} \right\rbrace^{5/3} \right] \right) & \cr
- {5 \over 2} \left({4 \over 3}\right)^{1/2} \left({\epsilon \over
\epsilon_{\rm sy, e}^{\rm br}} \right)^{1/2} \left(1 + {1 \over 5} \left[
{\gamma_{\rm e}^{\rm br} \over \Gamma} \right]^2 \right)
& if ${4 \over 3} \epsilon_{\rm sy, e}^{\rm max} \le \epsilon \le
{4 \over 3} (\gamma_{\rm e}^{\rm br})^2 \epsilon_{\rm sy, e}^{\rm br}$ \cr\cr
{11 \over 10} \, \left( {4 \over 3} \right)^{3/2} \, \left( {\epsilon \over
\epsilon_{\rm sy, e}^{\rm br}} \right)^{-1/2} \, (\gamma_{\rm e}^{\rm br})^2
& \cr
- {1 \over 2} \, \left( {4 \over 3} \right)^{1/2}\, \left( {\epsilon \over
\epsilon_{\rm sy, e}^{\rm br}} \right)^{1/2} \, \left( {\gamma_{\rm e}^{\rm br}
\over \Gamma} \right)^2 & \cr
- {3 \over 5} \, \left( {4 \over 3} \right)^{2/3}\, \left( {\epsilon \over
\epsilon_{\rm sy, e}^{\rm br}} \right)^{1/3} \, {(\gamma_{\rm e}^{\rm br})^2
\over \Gamma^{5/3}} & \cr
+ {1 \over 2} \, \left( {4 \over 3} \right)^{3/2}\, \left( {\epsilon \over
\epsilon_{\rm sy, e}^{\rm br}} \right)^{-1/2} \, (\gamma_{\rm e}^{\rm br})^2
\, \ln\left( {3 \epsilon \over 4 \epsilon_{\rm sy, e}^{\rm br} \, 
[\gamma_{\rm e}^{\rm br}]^2} \right) 
& if ${4 \over 3} (\gamma_{\rm e}^{\rm br})^2 \epsilon_{\rm sy, e}^{\rm br} 
\le \epsilon \le {4 \over 3} (\gamma_{\rm e}^{\rm br})^2 
\epsilon_{\rm sy, e}^{\rm max}$ \cr\cr
{1 \over 2} \, \left( {4 \over 3} \right)^{3/2}\, \left( {\epsilon \over
\epsilon_{\rm sy, e}^{\rm br}} \right)^{-1/2} \, (\gamma_{\rm e}^{\rm br})^2
\, \ln\left( {4 \epsilon_{\rm sy, e}^{\rm max} \, \Gamma^2 \over 3 \epsilon} 
\right) 
& if ${4 \over 3} (\gamma_{\rm e}^{\rm br})^2 \epsilon_{\rm sy, e}^{\rm max} 
\le \epsilon \le {4 \over 3} \Gamma^2 \epsilon_{\rm sy, e}^{\rm max}$ \cr\cr
}
\label{ssc_sp_II}
\end{equation}

The integrated luminosity in this spectrum is

\begin{equation}
L_{\rm SSC, e} = L_{\rm SSC}^0 \, N 
\label{Lsscnorm}
\end{equation}
where
\begin{equation}
N = \cases{ {1 \over 9} \, \left( {4 \over 3} \right)^{1/3} \, \Gamma^{17/3}
\, \left( {B \over B_{\rm cr}} \right)^{4/3}
& if $\gamma_{\rm e}^{\rm br} = \Gamma$ \cr\cr
\left\lbrace \left( {4 \over 3} \right)^2 \,
\epsilon_{\rm sy, e}^{\rm br} \, (\gamma_{\rm e}^{\rm br})^3 \right\rbrace
\, \Biggl\lbrace {5 \over 6} - {27 \over 10} \left( {\Gamma \over 
\gamma_{\rm e}^{\rm br}} \right)^{8/3} + {9 \over 20} \left( {\Gamma^3
\over [\gamma_{\rm e}^{\rm br}]^{11}} \right)^{1/3} & \cr
+ {5 \over 3} \left( {\Gamma \over [\gamma_{\rm e}^{\rm br}]^2} \right)^3 
+ {1 \over 3} {\Gamma \over (\gamma_{\rm e}^{\rm br})^4} - {31 \over 12} 
{\Gamma \over \gamma_{\rm e}^{\rm br}} + 2 \left( {\Gamma \over 
\gamma_{\rm e}^{\rm br}} \right)^2 \Biggr\rbrace 
& if $\gamma_{\rm e}^{\rm br} < \Gamma$ \cr}
\label{N_SSC}
\end{equation}

\newpage

\newpage

\newpage

\newpage

\newpage

\newpage

\begin{figure}
\plotone{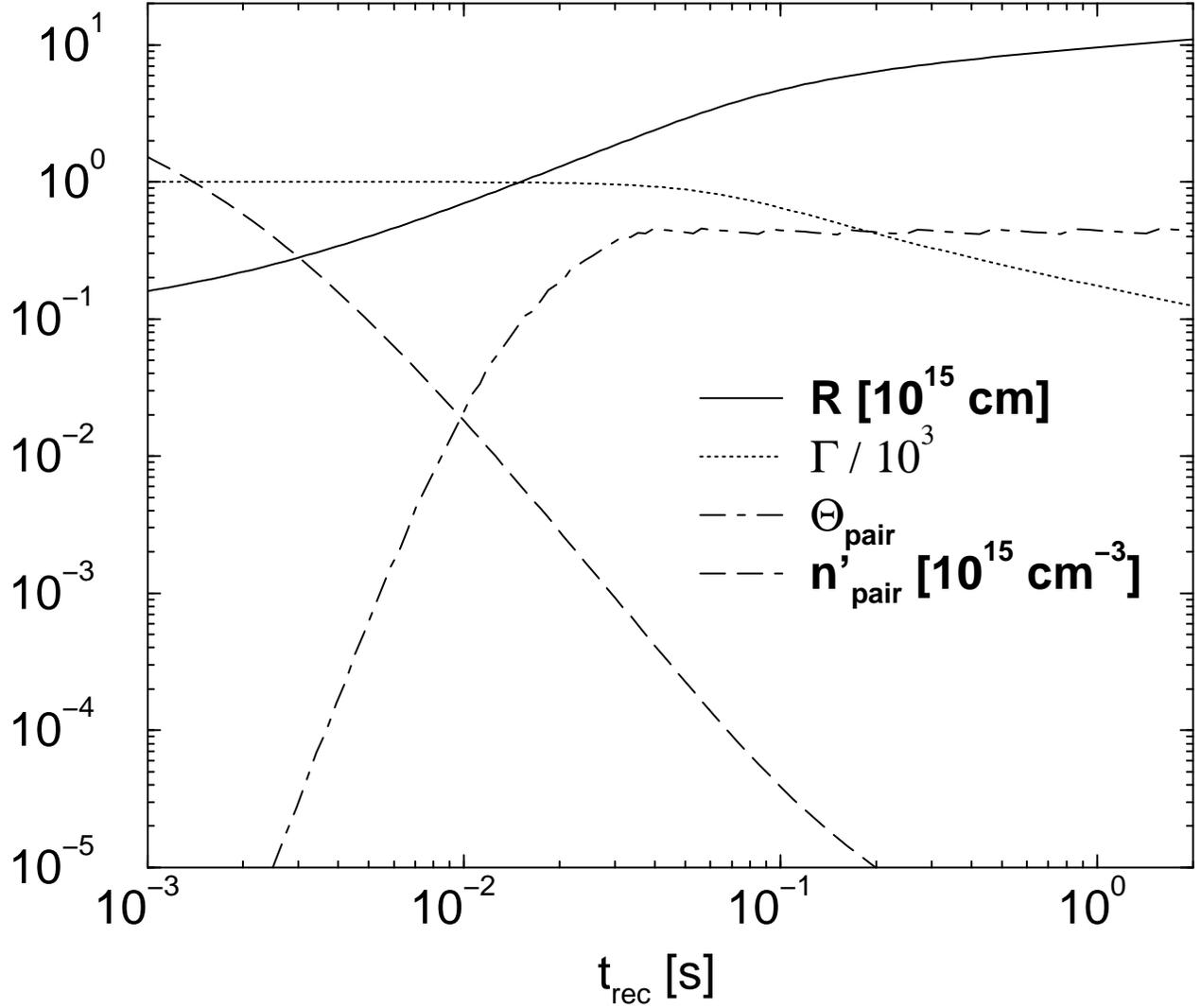}
\caption{Time evolution of several relevant quantities describing
the evolution of the plasmoid, for the same set of parameters as 
in Fig. \ref{composite2}.}
\label{temp2}
\end{figure}

\newpage

\end{document}